\documentclass[letterpaper, 10 pt, conference]{IEEEtran}
\IEEEoverridecommandlockouts
\usepackage{multicol}
\usepackage{multirow}

\usepackage{epsfig}
\usepackage{mathptmx}
\usepackage{times}
\usepackage{float}
\usepackage{color}
\usepackage{balance}
\usepackage{psfrag}
\usepackage{url}
\usepackage{stfloats}
\usepackage{amsmath}
\usepackage[amssymb]{SIunits}

\usepackage[utf8]{inputenc}
\usepackage{amssymb}
\usepackage{mathrsfs}
\usepackage{accents}

\usepackage{amsthm}

\usepackage{cuted}          %for one column block using \strip
\usepackage{algorithm}      %for algorithm
\usepackage{algpseudocode}  %for algorithm
\usepackage{ifsym}

\usepackage{graphicx}
\usepackage{hyperref}

\theoremstyle{definition}

\theoremstyle{definition}

\theoremstyle{plain}

\newcommand{\norm}[1]{\left\lVert#1\right\rVert}

\usepackage{xpatch}
\makeatletter
\xpatchcmd{\@thm}{\thm@headpunct{.}}{\thm@headpunct{}}{}{}
\makeatother
\xpatchcmd{\thm}{\@addpunct{.}}{\@addpunct{:}}{}{}

\makeatletter
\xpatchcmd{\@lemma}{\lemma@headpunct{.}}{\lemma@headpunct{}}{}{}
\makeatother

\hypersetup{
    colorlinks=true,
    linkcolor=blue,
    filecolor=blue,      
    urlcolor=blue,
    citecolor=blue,
}

\def\l{\left}
\def\r{\right}

\def\R{\mathbb{R}}

\title{\LARGE \bf Robust Target-relative Localization with Ultra-Wideband \\ Ranging and Communication}

\author{Thien-Minh Nguyen, Abdul Hanif Zaini, Chen Wang, Kexin Guo, Lihua Xie
\thanks{The authors are with School of Electrical and Electronic Engineering, Nanyang Technological University, Singapore 639798, 50 Nanyang Avenue.}
\thanks{The research is partially supported by the ST Engineering - NTU Corporate Lab through the NRF corporate lab@university scheme.}
\thanks{Corresponding author's email: elhxie@e.ntu.edu.sg.}
\thanks{First author's email: e150040@e.ntu.edu.sg, thienminh.npn@gmail.com}
}

\begin{document}

\maketitle

\begin{abstract}
In this paper we propose a method to achieve relative positioning and tracking of a target by a quadcopter using Ultra-wideband (UWB) ranging sensors, which are strategically installed to help retrieve both relative position and bearing between the quadcopter and target. To achieve robust localization for autonomous flight even with uncertainty in the speed of the target, two main features are developed. First, an estimator based on Extended Kalman Filter (EKF) is developed to fuse UWB ranging measurements with data from onboard sensors including inertial measurement unit (IMU), altimeters and optical flow. Second, to properly handle the coupling of the target's orientation with the range measurements, UWB based communication capability is utilized to transfer the target's orientation to the quadcopter. Experiment results demonstrate the ability of the quadcopter to control its position relative to the target autonomously in both cases when the target is static and moving.
\end{abstract}

\section{Introduction} \label{sec:intro}

In multi-robot systems, it is usually expected that each robot be able to determine its relative position to other robots to carry out tasks such as UGV-MAV (Unmanned Ground Vehicle - Micro Aerial Vehicle) cooperation or movement in formation. A common approach in such and similar scenarios is to assume that each robot can determine its position in a globally-shared frame and transmit this information to its neighbors. However, satellite-based systems such as the GPS (Global Positioning System) are limited to open and uncluttered outdoor environments. Alternative methods, such as motion tracking camera systems or radio-based positioning systems, are dependent on careful setup in the area of operation. Examples of the latter include our previous work on UWB-based localization \cite{guo2016ultra, nguyenultra, wang2017ultra} and similar work by others in \cite{mueller2015fusing, ledergerber2015robot}.

Alternatively, some works assume some proximity sensing capability. One approach uses distance sensing methods such as with infrared distance sensors \cite{pickem2015gritsbot} or laser scanners \cite{zakhar2015distributed}. However, this approach does not provide identification of neighbors and is relatively short-ranged. Another highly promising method which has been a popular topic of research is the use of computer vision systems. The vision-based method presents a favorable alternative to the system presented in this paper but it is not without limitations. See e.g. \cite{conroy20143, pestana2014computer, pestana2013vision}. Briefly, the limitations include limited field-of-view, short range, and possibly demanding computing power capability, which lightweight aerial robots cannot cater for.

\begin{figure}[h]
\centering
    \includegraphics[width=3.25in]{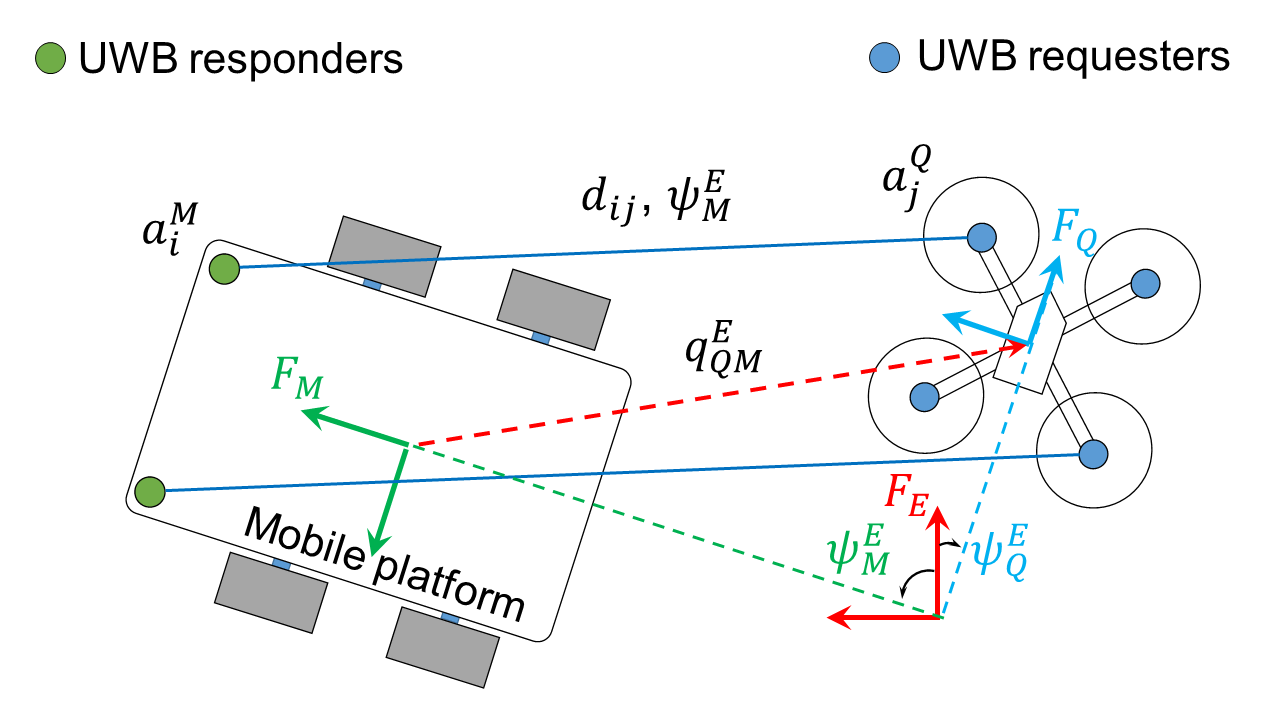}
    \caption{System overview. In this work, two ranging sensors, each has two separate antennae, are installed on the MAV. Thus, we can effectively count up to four UWB requester nodes. By using different channels, two range measurements can be acquired simultaneously by the two ranging sensors on the MAV. A ranging pattern is pre-programmed on the nodes to cycle through eight useful measurements in four consecutive steps. Note that in $q^{E}_{QM}$, the subscript $(\cdot)_{QM}$ is used to indicate that the direction of vector $q$ is from the origin of the frame $F_M$ to the origin of the frame $F_Q$, and the subscript $E$ is used to indicate that the coordinates are in reference to the frame $F_E$.}
    \label{fig: sysarch}
\end{figure}

In this paper, we present a system for single target relative localization by a single mobile robot. Here we utilize UWB-based TWTOF (Two Way Time of Flight) asynchronous ranging measurements with the target mobile robot to determine relative position without the need for additional external devices. The position estimation method is similar to our previous works in \cite{guo2017ultra, nguyenultra}. Instead of multiple anchors placed strategically in the operation area, both robots carry multiple UWB antennae placed in suitable configurations. Moreover, we develop the system for the MAV-UGV cooperation scenario (Fig. \ref{fig: sysarch}), and integrate IMU, altimeter and optical correlation flow data in an Extended Kalman Filter, to provide relative position data that is sufficiently accurate and stable for feedback-controlled flight. This ranging-based method is omnidirectional and allows other sensors such as cameras to be used for other tasks instead of keeping it trained on the target \cite{gui2013airborne}. Using this system, the MAV would be capable of performing tasks with accurate positioning around the UGV and dock precisely with it. For such application, the UGV can be assumed to stay fixed, however in this work we focus on experimenting the robustness of the system when the target is moving and the MAV does not know its velocity.

Several relevant works can be found in recent literature. In \cite{fidan2013adaptive}, Fidan et al. addressed the problem of simultaneous localization and tracking of a moving target using distance measurement in a rigorous mathematical framework. The static target case was then reassessed with discrete time in \cite{fidan2015adaptive}. Their work tackles the challenging single source distance measurement scenario but still requires reliable self-position information. Moreover, only simulations with a single-integrator model were provided. In another notable work with physical implementation \cite{fabresse2014localization}, Fabresse et al. integrate vision and range measurements to map the positions of radio nodes, but with the aerial robot’s position already known. In their experiments, visual markers were used for ground truth. In \cite{Hepp2016}, Hepp et al. use multiple UWB antennae, similar to our implementation, to estimate the position of the target using iterated EKF. However, flight experiments were still reliant on a motion capture system for self-localization leaving relative position feedback controlled flight to future work. UWB transceivers, as mentioned earlier, have also been used in GPS replacement systems such as in \cite{mueller2015fusing, ledergerber2015robot, guo2017ultra, nguyenultra} with variations in model and use of TOA (Time of Arrival) or TWTOF.

Our main contribution is a complete relative localization system integrating UWB ranging measurements with other standard MAV sensors such as IMU, altimeters and a computationally efficient optical flow so-called \textit{correlation flow}. Unlike the works in \cite{fabresse2014localization, Hepp2016}, our system provides relative position estimates that are sufficiently accurate and stable for feedback-controlled flight and are independent of external systems for localization such as GPS or camera-based motion tracking. Additionally, we exploit the UWB communication capability to share information between robots.

The remaining of this paper is organized into two main parts. Section \ref{sec:sysarch} details the basic components of the system, most notably the sensor models and some pragmatic techniques in our EKF design. Section \ref{sec:exp} presents the main achievements of our approach with two sets of experiments. The first set demonstrates the ability of the system for omni-directional positioning and the second set demonstrates the robustness of the estimation on the MAV over the unknown target's odometry. As relative position is our main goal, root mean square error (RMSE) and standard deviation (SD) of the relative position estimates are calculated and reported in detail, however other data such as orientation and velocity estimate are also presented depending on the context. Finally, we conclude and discuss the potential for future development in Section \ref{sec:conclusion}.

%%%%%%%%%%%%%%%%%%%%%%%%%%%%%%%%%%%%%%%%%%%%%%%%%%%%%%%%%%%%%%

\section{System description} \label{sec:sysarch}

\subsection{Problem Formulation}

Our main goal in this work is to estimate the quadcopter's position relative to the mobile platform, defined as $q^E_{QM}$, using the main sources of information as illustrated Fig. \ref{fig: sysarch} (notice the caption for some notational implications). First, multiple UWB nodes are installed on the quadcopter MAV and the mobile platform which can be a manned or unmanned vehicle. For convenience, we call this mobile platform the \textit{target}. The UWB nodes on the target are called the responders and the UWB nodes on the quadcopter are named requesters. We denote the location of a responder $i$ in the frame $F_Q$ fixed on the quadcopter as $a_i^Q$. Similarly $a_j^M$ is the location of a requester in the frame $F_M$ attached to the target and the relative position between the MAV and the target is denoted as $q^E_{QM}$. The MAV can measure the distance $d_{ij}$ between a requester $i$ and a responder $j$ via the TWTOF protocol. This measurement is the most important source of observation for localization in our system.

Besides the relative position, in close proximity the orientations of the MAV and target are also critical. In this work, we assume both MAV and target can estimate their orientation relative to an inertial frame of reference $F_E$. The MAV can receive the measurement of the target's orientation $\psi^E_M$ relative to the inertial frame $F_E$ via the responding messages $M_{rspd}$ in the TWTOF transactions (Fig. \ref{fig:twof}).

It can be seen in Fig. \ref{fig: sysarch} that if all of the eight distance measurements $d_{ij}$, the mobile platform's orientation $\psi^E_M$, and the MAV's altitude are obtained at the same time, then we can directly calculate the relative position and orientation of the MAV to the target. However, since our ranging method is based on TWTOF, each UWB node can only make response/request to one other node in each transaction. Thus, the quadcopter can only obtain as many range measurements at a time as there are many requesters. In fact, in our system the MAV can only acquire at most two range measurements simultaneously even though four requester nodes are illustrated in Fig. \ref{fig: sysarch}. This is because we can only install two ranging sensors on the MAV and the number of requester nodes are extended by carrying out ranging measurement over spatially separated antennae. Besides unsynchronized observations, measurement noise would also affect the accuracy of the estimate. Thus, the MAV has to carry out some prediction between the arrivals of new measurements and fuse these sources of information asynchronously to obtain a robust estimate for autonomous flight. In our case an EKF is developed for this purpose. The state vector of the quadcopter under this EKF approach is chosen as follows:
\begin{equation} \label{equ:statevector}
    X \triangleq \mathrm{col}\left\{\psi,\ {v},\ {q}\right\},
\end{equation}
where the states are defined as follows:
\begin{itemize}
    \item ${\psi} = \left[{\psi}_w,\ {\psi}_x,\ {\psi}_y,\ {\psi}_z\right]'$
    is the unit quaternion representing the quadcopter's orientation relative to the frame $F_E$, i.e. $\psi^E_Q$ in Fig. \ref{fig: sysarch}. Here we use the convention where ${\psi}_w$ is the real part and $\left[{\psi}_x, {\psi}_y, {\psi}_z\right]'$ is the imaginary part.
    
    \item ${v} = \left[{v}_x,\ {v}_y,\ {v}_z\right]'$ is the relative velocity between the quadcopter and the target in the frame $F_E$, i.e. $v^E_{QM}$. Here $v_x$, $v_y$, $v_z$ are the Cartesian coordinates of the velocity in the frame $F_E$,
    
    \item ${q} = \left[{q}_x,\ {q}_y,\ {q}_z\right]'$ is the quadcopter's position relative to the target, referenced in the frame $F_E$, i.e $q^E_{QM}$. Here ${q}_x$, ${q}_y$, ${q}_z$ are the Cartesian coordinates of ${q}$ in the frame $F_E$.
\end{itemize}

Note that ${X} \triangleq \mathrm{col}\l\{{\psi}^E_Q,\ {v}^E_{QM},\ {q}^E_{QM}\r\}$ could be a more expressive notation for the state vector, yet as the implications of the subscripts and superscripts can be easily inferred from the context, we opt to omit these extra notations to keep the notation concise.
With the state vector defined in (\ref{equ:statevector}), under the EKF paradigm, we can define a state estimate vector $\hat{X} = \mathrm{col}\left\{\hat{\psi},\ \hat{v},\ \hat{q}\right\}$ of the state vector $X$. Hence, $\hat{X}$ can be updated using the observations $d_{ij}$, $\psi^E_M$, $a_i^M$, $a_j^Q$ whose relationship with $X$ is described in the next part. As these measurements are obtained at different rates, an EKF is suitable to fuse all of these observations with other onboard sensors in an asynchronous fusion scheme to robustly estimate the state vector $X$ so that feedback-control flight is sustainable.

\subsection{Sensors}

In this part we describe the model of sensors used in our system and discuss some of the important pragmatic measures to successfully achieve a robust estimation of the system states.

\underline{\textbf{Inertial Measurement Unit:}}IMU is used to mainly estimate the orientation. We denote the data obtained from a IMU as follows:
    \begin{itemize}
        \item $\omega \in \R^3$ is the angular rate from the gyroscope.
        \item $a \in \R^3$ is the acceleration from the accelerometer.
        \item $m \in \R^3$ is the earth's magnetic field from the magnetometer.
    \end{itemize}

All of the aforementioned sensor data are in the quadcopter's body frame $F_Q$. The relationship between $\omega$, $a$ and $m$ with the state vector is described in following differential equations derived from the dynamics of a \textit{strapdown inertial navigation system}:
\begin{align}
    \dot{\psi} &= \frac{1}{2} (\psi \circ \psi_\omega), \label{equ:strapdownq}\\
    \dot{v} &= R^E_Q(\psi) a - \textbf{g} - \dot{v}^E_M, \label{equ:strapdownv}\\
    \dot{q} &= v, \label{equ:strapdownp}\\
    m &= R^Q_E(\psi)\textbf{m},\label{equ:strapdownm}
\end{align}
where $(\circ)$ denotes the quaternion multiplication. $\psi_\omega$ is the quaternion with zero in the real component and the angular rate $\omega$ in the imaginary part, $R^E_Q(\psi)$ is the rotation matrix constructed from the quaternion $\psi$ and $R^Q_E(\psi)$ is its inverse. $\textbf{g}$ is the Earth's gravity. $v^E_M$ is the velocity induced by the translation and rotation of the frame $F_M$. We assume these motions are small enough so that $\dot{v}^M_E$ can be considered as a process noise. Finally $\textbf{m}$ is the direction of the Earth's magnetic field in the inertial frame $F_E$.

Notice that equations (\ref{equ:strapdownq}), (\ref{equ:strapdownv}), (\ref{equ:strapdownp}) are used for prediction while (\ref{equ:strapdownm}) is for the correction stage. Moreover, the gyroscope biases in the IMU measurements are also accounted for by introducing extra states beside the ten main states in (\ref{fig: sysarch}). We will update the state estimate $\hat{X}$ of $X$ using the discretized and linearized versions of the differential equations \eqref{equ:strapdownq}, \eqref{equ:strapdownv}, \eqref{equ:strapdownp}. These tasks follow quite well-established procedures in implementation of strapdown inertial navigation system, thus the details are omitted for the sake of brevity.

\begin{figure}[h]
    \centering
    \includegraphics[width=0.4\textwidth]{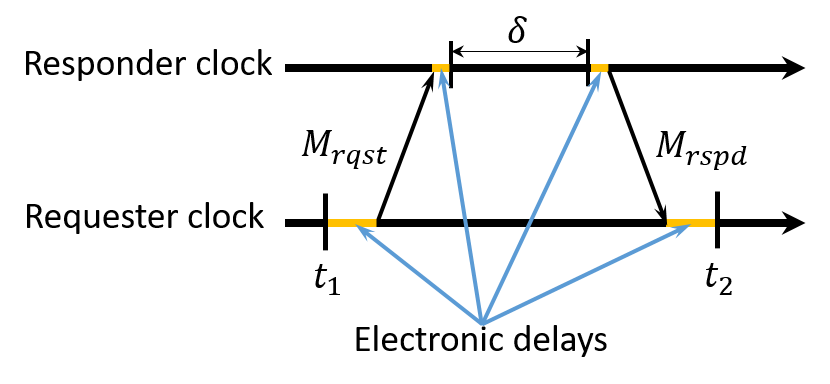}
    \caption{A diagram of TWTOF ranging protocol.}
    \label{fig:twof}
\end{figure}

\underline{\textbf{Orientation-coupled range measurements:}}Fig. \ref{fig:twof} illustrates a TWTOF transaction in our system. As can be seen in this diagram, the distance between the requester and responder nodes can be calculated using the TWTOF model as follows:
\begin{equation}
    d = c\frac{t_2-t_1 - \delta }{2} - \Delta,
\end{equation}
where $c$ is the speed of light, $t_1$ and $t_2$ are the time instances when the request and responses messages are recorded on the requester's clock respectively, $\delta$ is a predefined period that the responder has to wait before responding, $\Delta$ is the gross distance bias due to electronic delays in the extension cables and connectors. This constant $\Delta$ has to be measured empirically for each pair of a requester and a responder.

We can now state the relationship between the UWB measurement and our selected state vector $X$. Denote $||.||$ as the Euclidean norm of a vector in $\mathbb{R}^3$, the relationship between our measurement and state vector can be stated as follows:
\begin{equation}
    d_{ij} = \norm{q + R^E_Q(\psi) a^Q_i - R^E_M(\psi^E_M) a^M_j}\label{equ:dist},
\end{equation}
 where $ R^E_Q(\psi)$ is the direct cosine matrix (DCM) constructed from the quarternion states $\psi$, and $R^E_M(\psi^E_M)$ is the DCM constructed from the orientation representation of the target's frame $F_M$ relative to $F_E$.
 
 We can see that the distance measurement couples both the position and orientation of the MAV. Thus, one advantage of having multiple requester nodes and at least two responder nodes in Fig. \ref{fig: sysarch} is that we can disambiguate the orientation of the MAV from the distance observation. Otherwise the yaw can very often drift if there is only a single UWB requester node on the MAV or a single requester node on the target.

\underline{\textbf{Optical flow:}}
In this work, we employ the recently proposed optical flow algorithm, \textit{kernel cross-correlator}-based \textit{correlation flow} \cite{wang2017cross}, to obtain an accurate velocity estimation.
It is open source\footnote{\url{https://github.com/wang-chen/correlation_flow}}, computationally efficient and robust to motion blur. The key feature of our method is the use of kernel cross-correlator to efficiently predict the transformation in Fourier domain between the current and previous image, including translation, rotation, and scale.
After this operation, the position of the highest value in the correlation output will identify the most suitable translation, rotation and vertical movements of the camera between two frames. 
In this work we only use a simple model of 2D translational optical flow with the following measurement model:
\begin{equation}\label{equ:flow}
\left[\begin{array}{c} v_{f_x} \\v_{f_y}\end{array}\right] = \left[\begin{array}{ccc} 1 &0 &0 \\ 0 &1 &0\end{array}\right]\frac{1}{q_z}R^Q_E(\psi)v.
\end{equation}

The current implementation assumes that the camera's image plane is always parallel with the ground plane.
Therefore the use of $R^Q_E(\psi)$ in the model (\ref{equ:flow}) is not an exact description. However, as the MAV's angles are given some threshold (which is to limit the maximum speed), the effect of roll and pitch is minor and can be lumped to the process noise. This approach has been validated in actual autonomous flight tests with only one camera, IMU and onboard altimeters. The video recording of this test can be viewed online\footnote{\url{https://youtu.be/DEjwjzJX3b4}}.

\underline{\textbf{Altimeters}}
For altitude estimation a laser range finder and barometer data are used to measure the distance from the MAV to the floor at an angle. The relationship between the laser range finder reading $l$ and the state vector can be stated as follows:
\begin{equation}
    l = \frac{q_z}{\psi_w^2 - \psi_x^2 - \psi_y^2 + \psi_z^2}, \label{equ:lidar}
\end{equation}
where $q_z$ is the altitude of the MAV in the frame $F_E$ and the denominator on the right hand side of (\ref{equ:lidar}) is the cosine of the angle between the vectors with coordinates $\left(0,\ 0,\ 1\right)$ in both $F_E$ and $F_Q$ frames.

The barometer reading $b$ is directly related to the altitude $q_z$ with an offset $b_0$:
\begin{equation}
    b = q_z + b_0,
\end{equation}

\subsection{Asynchronous fusion}
In this section we describe the workflow to fuse multiple sensor data described in previous parts. Algorithm \ref{algo:ekfseqfusion} summarizes the main operations of this fusion thread.

\begin{algorithm}
        \caption{EKF - Asynchronous fusion}
        \begin{algorithmic}[1]
        
        \While {Thread is healthy}
            
            \State \textbf{Poll\_sensor\_data}()
            
            \If {New sensor data available}
                \If{New gyroscope and accelerometer data} \label{algo:fuseimu}
                    \State  \textbf{Predict\_orientation\_velocity}($\omega$, $a$)
                \EndIf
                
                \If{New magnetometer data} \label{algo:fusemag}
                    \State  \textbf{Fuse\_magnetometer}($m$)
                \EndIf
                
                \If{New UWB data} \label{algo:fusedist}
                    \State  \textbf{Fuse\_UWB\_distance}($d_{ij}$, $a^Q_i$, $a^E_j$)
                \EndIf
                
                \If{New correlation flow data} \label{algo:fuseflow}
                    \State  $v_{f_z} = [0, 0, 1]' R^Q_E(\hat{\psi}) \hat{v}/q_z$ \label{algo:artiflow}
                    \State  $v = q_z R^E_Q(\hat{\psi})\left[v_{f_x}, v_{f_y}, v_{f_z}\right]'$ \label{algo:augflow}
                    \State  \textbf{Fuse\_velocity}($v$)
                \EndIf
                
                \If{New laser range finder data}
                    \State  $h = l(\hat{\psi}_w^2 - \hat{\psi}_x^2 - \hat{\psi}_y^2 + \hat{\psi}_z^2)$ \label{algo:findaltitude}
                    \State  \textbf{Fuse\_altitude}($h$)
                \EndIf \label{algo:fuselidar}
                
                \If{New barometer data}
                    \State  $h = b - b_0$
                    \State  \textbf{Fuse\_altitude}($h$)
                \EndIf \label{algo:fusebaro}
            \EndIf
        \EndWhile
        \end{algorithmic} \label{algo:ekfseqfusion}
\end{algorithm}

As the sensors are managed by different threads with different update rates, the data from these sensors are not synchronized. Hence, in our fusion scheme, a fusion thread will keep polling for new sensor data and corresponding stages of prediction/update will be selected according to the sensor type.
Note that in steps \ref{algo:fuseimu} to \ref{algo:fusebaro} of Algorithm \ref{algo:ekfseqfusion}, discretized versions of equations (\ref{equ:strapdownq}), (\ref{equ:strapdownv}), (\ref{equ:strapdownp}), (\ref{equ:strapdownm}), (\ref{equ:dist}) and (\ref{equ:flow}) are linearized to predict/update the state estimate and the error covariance matrix in a canonical way. However in steps $\ref{algo:fuseflow}$ to $\ref{algo:fuselidar}$, we pragmatically ignore the mathematical coupling of the observation with the orientation states. Specifically, for the optical flow estimate, we first convert the current velocity estimate to the body frame and scale it by the MAV's altitude estimate to obtain a so-called \textit{artificial flow} (step \ref{algo:artiflow}). The $z$ component of this \textit{artificial flow} is thus extracted and combined with the 2D optical flow data from camera to form a 3D \textit{augmented flow} measurement (step \ref{algo:augflow}). This \textit{augmented flow} vector is then used to produce an approximation of velocity measurement in the inertial frame $F_E$ by multiplying it with $R^E_Q(\hat{\psi})$ and the altitude estimate. This approach helps reduce unnecessary computation given that the MAV's attitude is not supposed to vary too much and any error can be lumped to the process noise. The same rationale is applied to the treatment of the laser range finder for the altitude observation in step \ref{algo:findaltitude}. The success of our flight tests would validate this pragmatism in retrospect.

\section{Experimental Results} \label{sec:exp}

In this part we will describe in detail the physical implementation of our system. Video recording of our experiments can be found at \url{https://youtu.be/5zelvj_xPzM}. The datasets collected from these experiments can be downloaded at \url{https://github.com/BritskNguyen/icra2018_uwb_sensor_fusion}.

\begin{figure}
    \centering
    \includegraphics[width=0.45\textwidth]{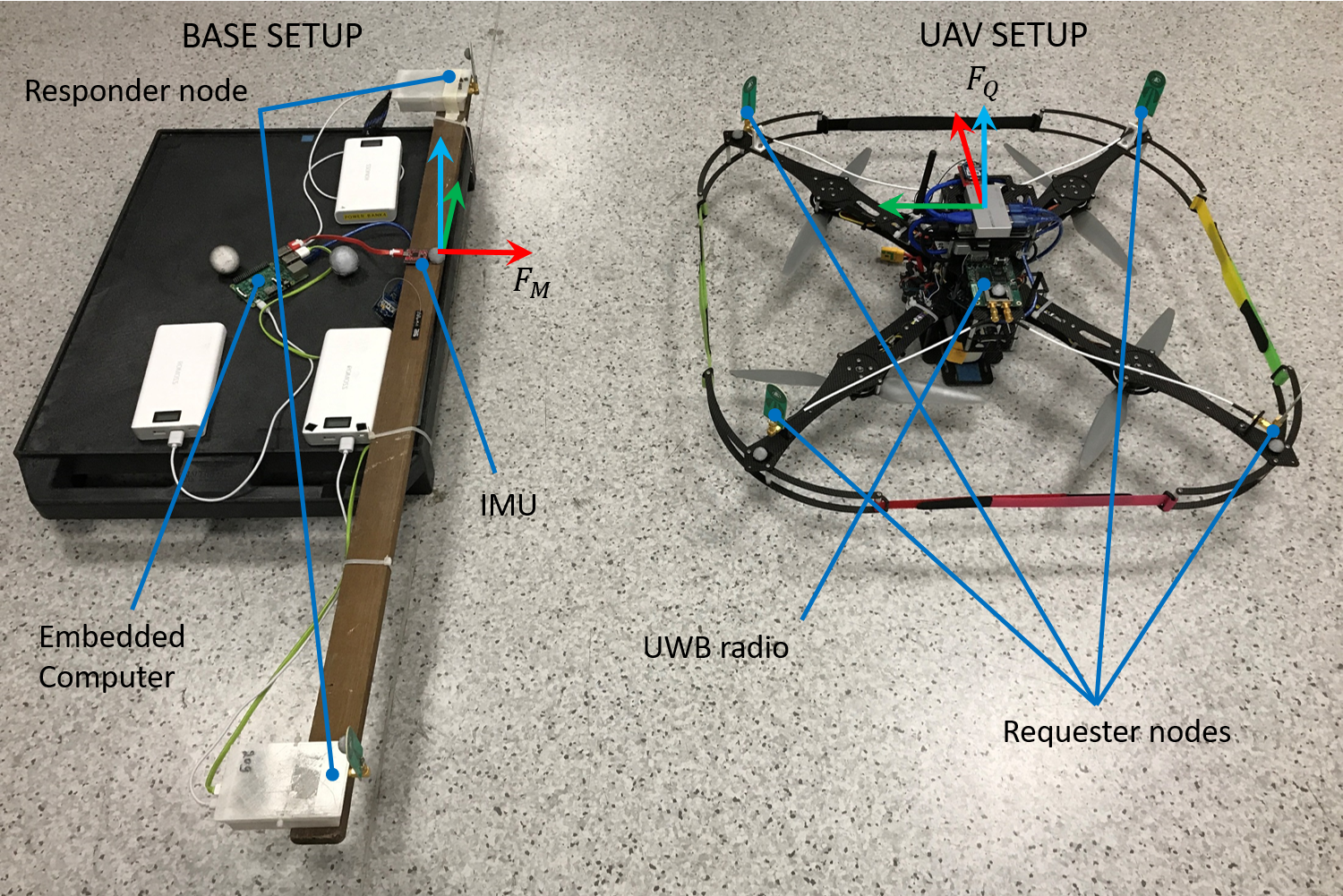}
    \caption{Actual equipment used in the experiments.}
    \label{fig:setup}
\end{figure}

Fig. \ref{fig:setup} shows the main components in our experiments. On the target's side, two UWB radios are used as responders in ranging and communication transactions (we notice that a ranging error below $2cm$ is reported by the manufacturer for the latest version\footnote{\url{http://www.timedomain.com/products/pulson-440/}}). These radios are hosted by a small-size embedded computer whose main job is to query the orientation data from a low-cost IMU module comonly used in robotics research\footnote{\url{http://wiki.ros.org/myahrs_driver}}. This IMU data is then relayed to the UWB radio's buffer to be included in the response messages of the TWTOF ranging transaction.

\begin{figure*}
        \centering
        \includegraphics[width=0.475\textwidth]{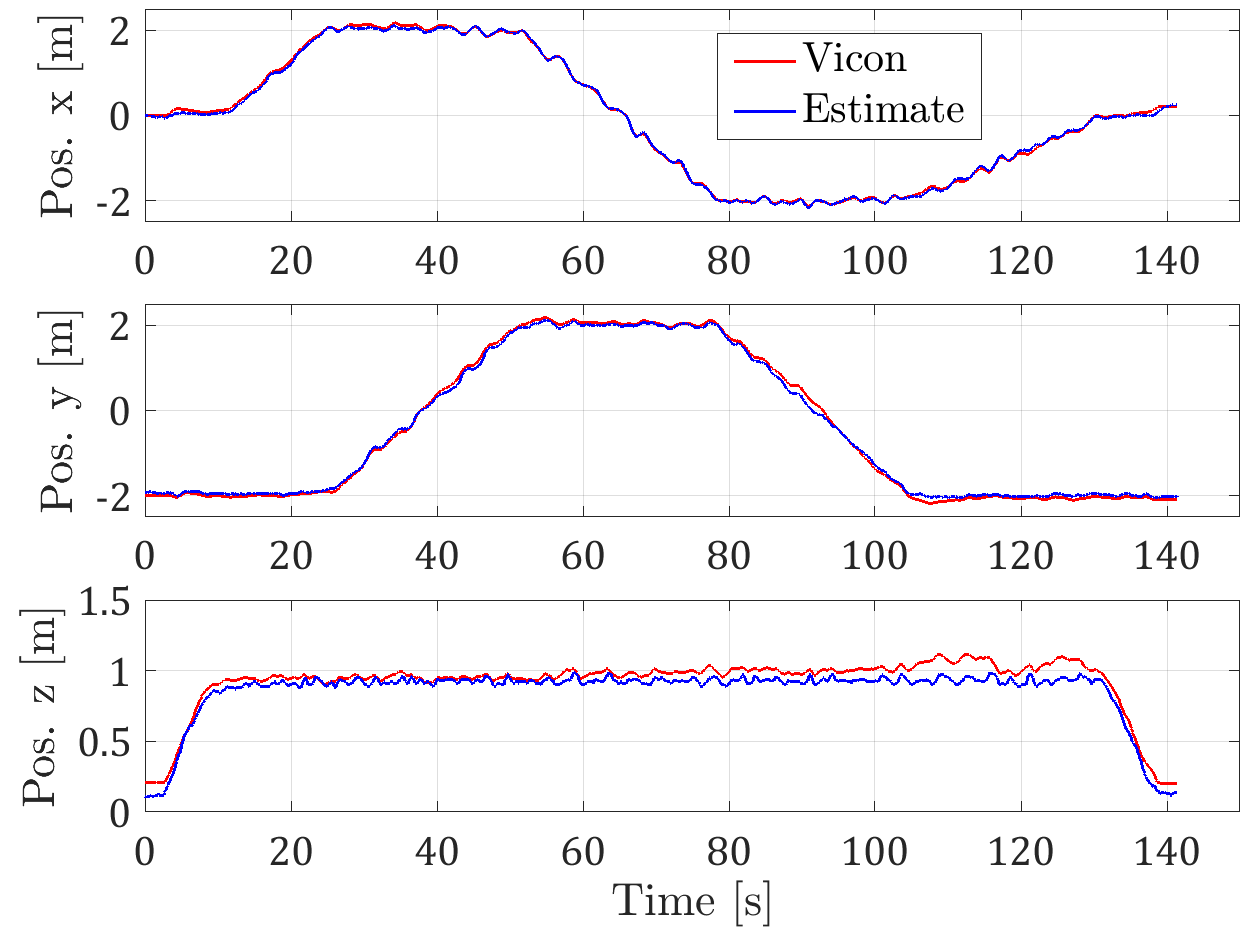}
        \includegraphics[width=0.475\textwidth]{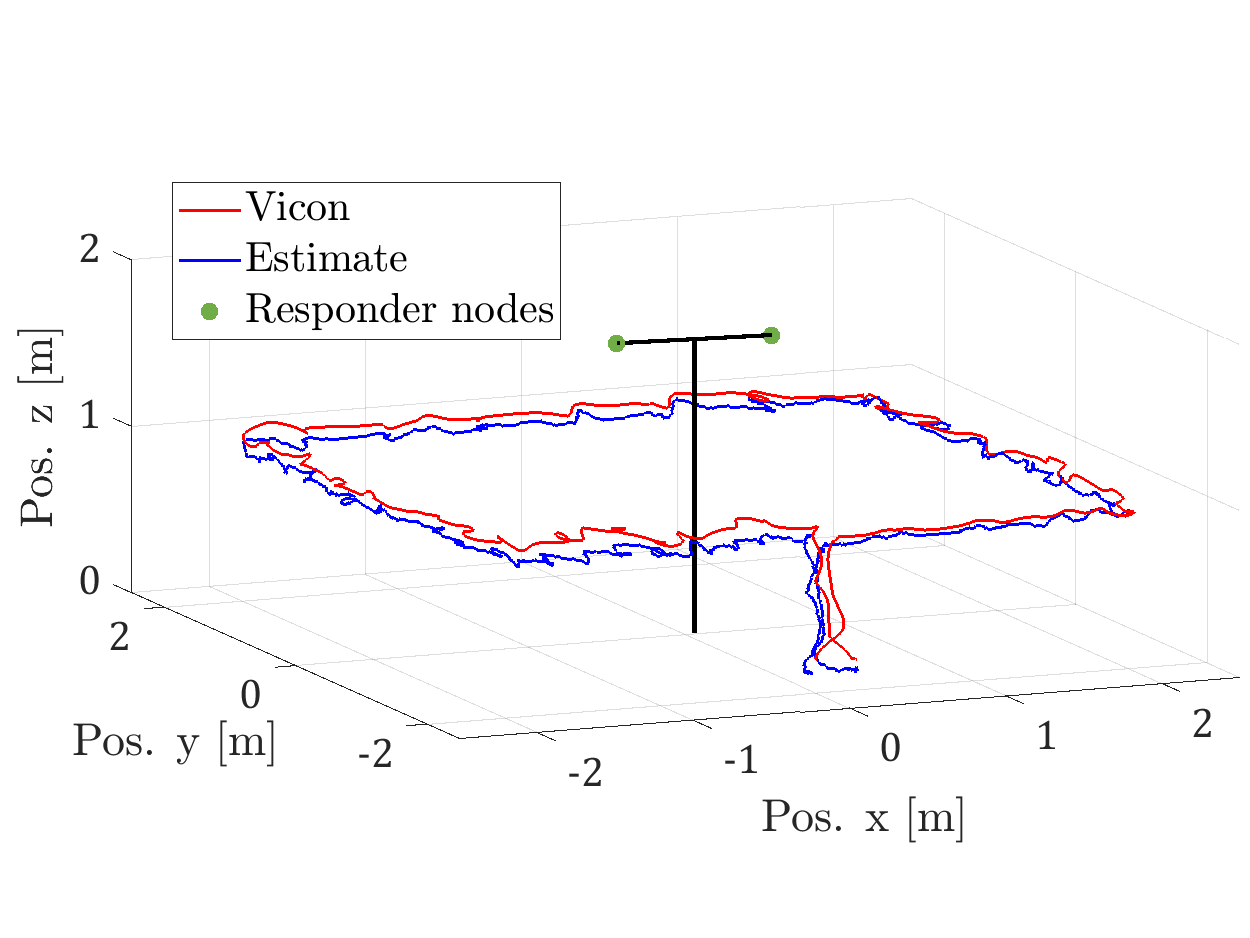}

        \centering
        \includegraphics[width=0.475\textwidth]{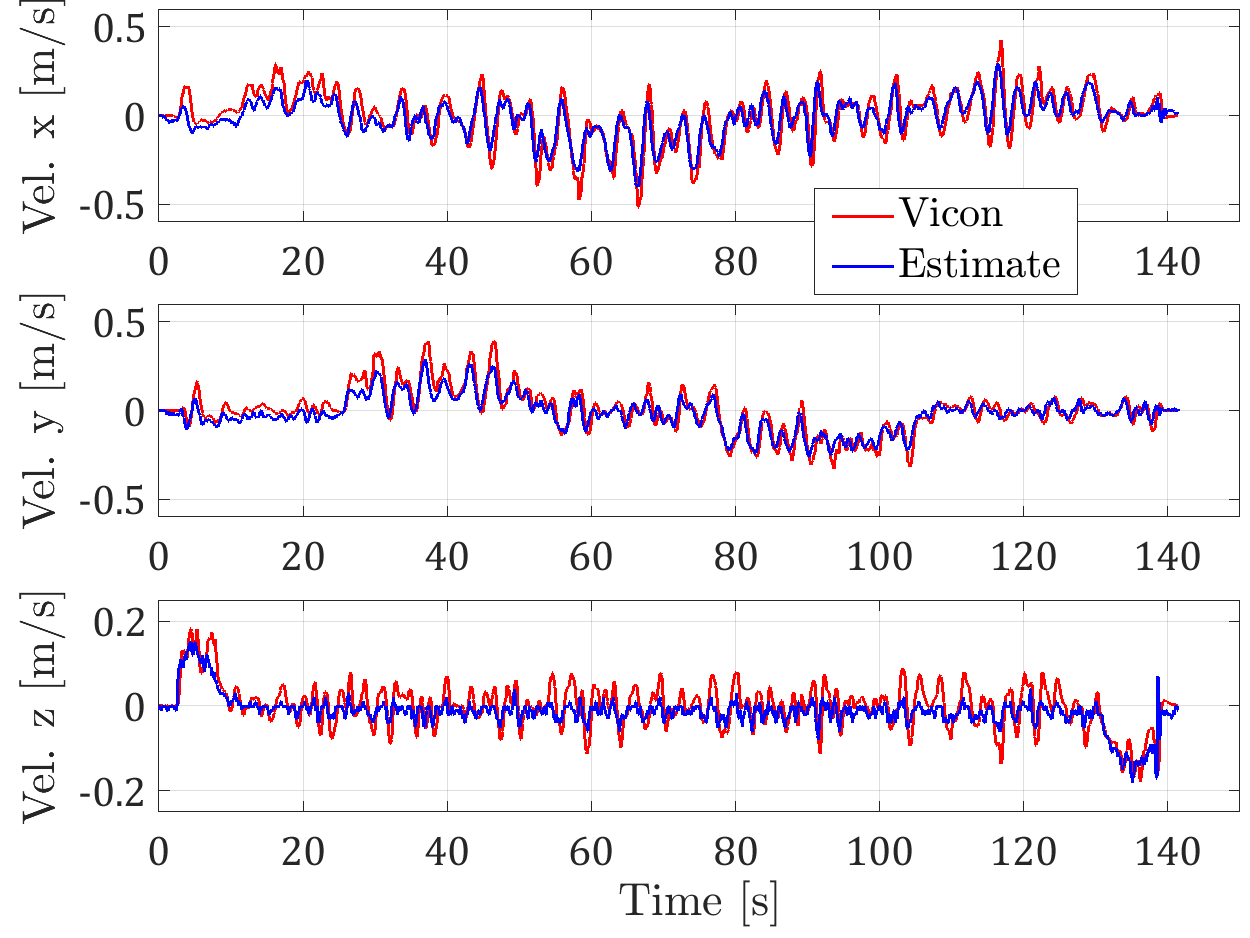}
        \includegraphics[width=0.475\textwidth]{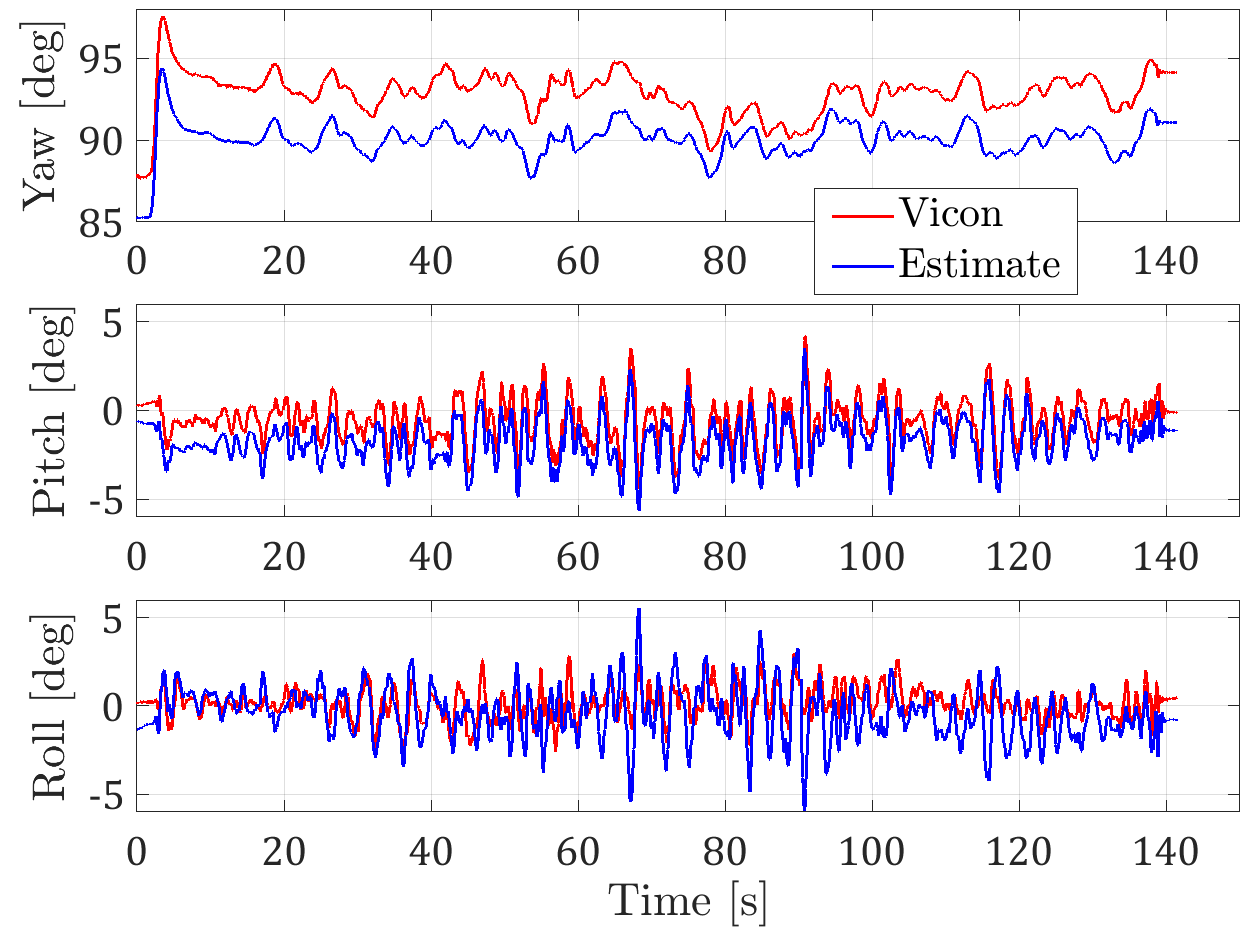}
        
\caption{Static target experiment: The MAV follows a predefined path of a $4m \times 4m$ square at $0.9m$ altitude around the two static responders. All data are in reference to the frame $F_E$. The responders' coordinates in the frame $F_M$ are $\left(0.04,\ -0.57,\ 1.753 \right)$ and $\left(0.035,\ 0.424,\ 1.778\right)$, which means the two responder nodes are only separated for approximately 1 meter apart.}
\label{fig:nearanc}
\end{figure*}

% \begin{figure*}
%         \centering
%         \includegraphics[width=0.475\textwidth]{exp/path3Dsq_near.png}
%         \includegraphics[width=0.475\textwidth]{exp/pos_sq_near.png}

%         \centering
%         \includegraphics[width=0.475\textwidth]{exp/vel_sq_near.png}
%         \includegraphics[width=0.475\textwidth]{exp/ang_sq_near.png}
        
% \caption{Static target experiment: The MAV follows a predefined path of a $4m \times 4m$ square at $0.9m$ altitude around the two static responders. All data are in reference to the frame $F_E$. The responders' coordinates in the frame $F_M$ are $\left(0.04,\ -0.57,\ 1.753 \right)$ and $\left(0.035,\ 0.424,\ 1.778\right)$, which means the two repsonder nodes are only separated for approximately 1 meter apart.}
% \label{fig:nearanc}
% \end{figure*}

\begin{figure*}[h]
    \centering
        \includegraphics[width=0.325\textwidth]{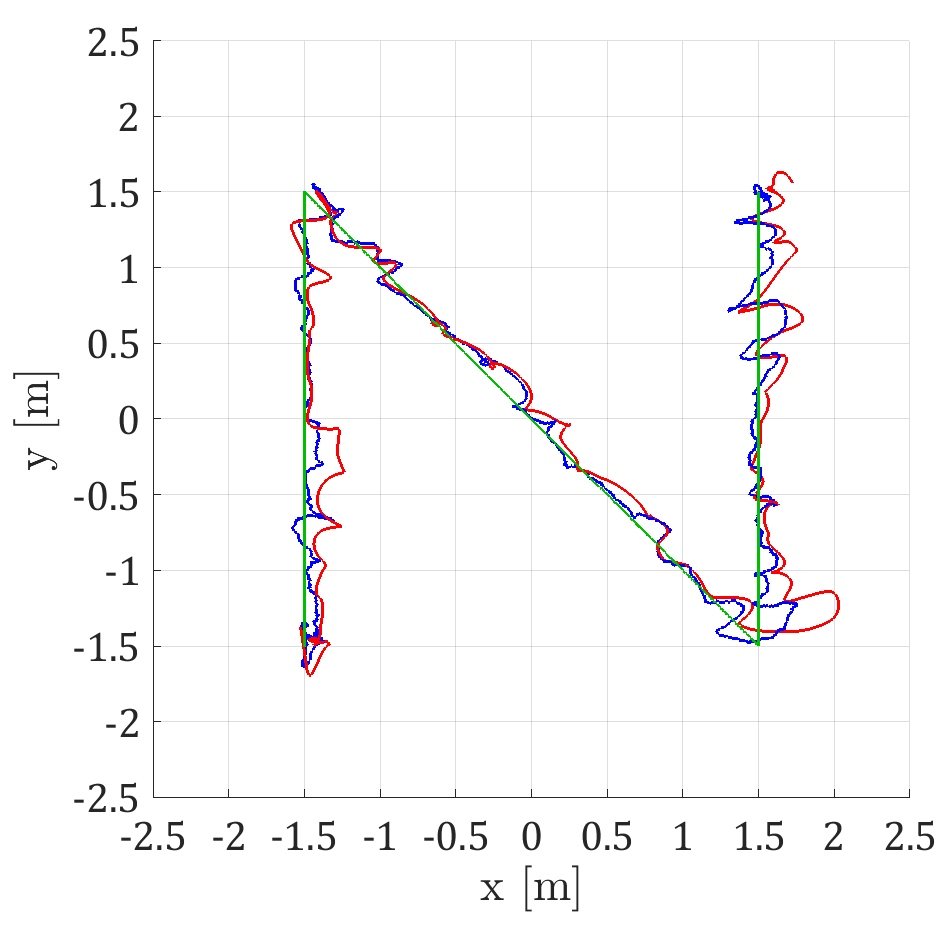}
        \includegraphics[width=0.325\textwidth]{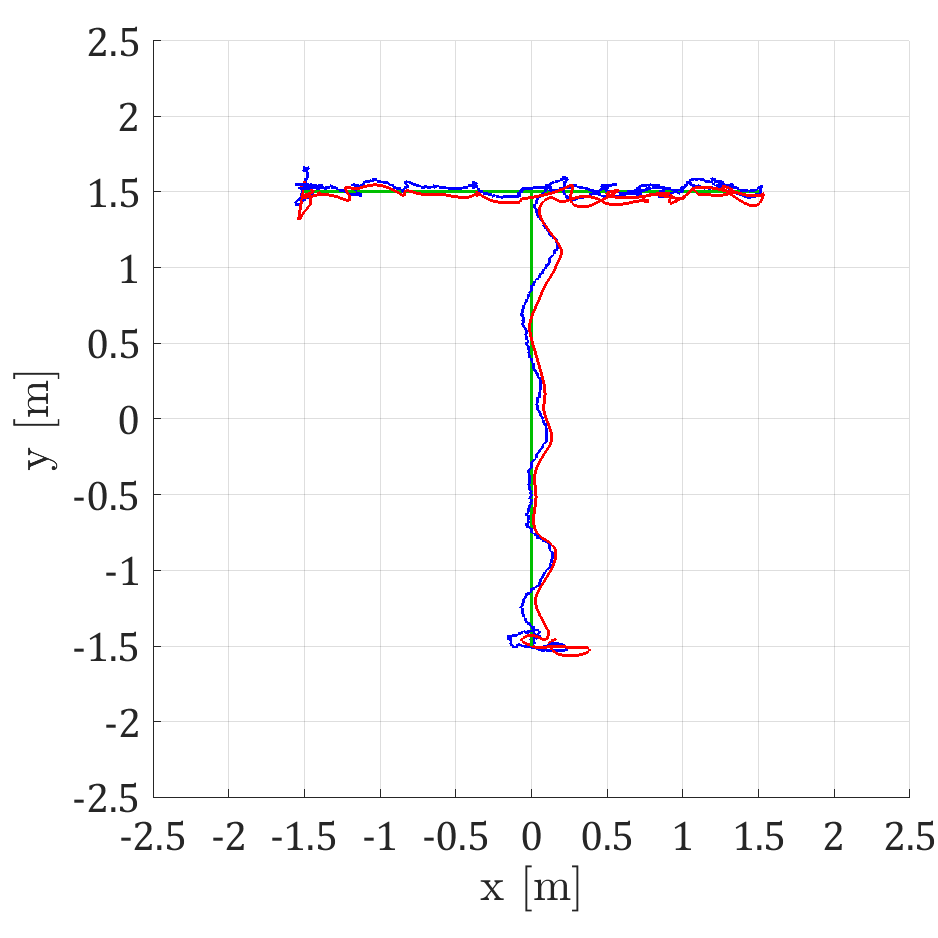}
        \includegraphics[width=0.325\textwidth]{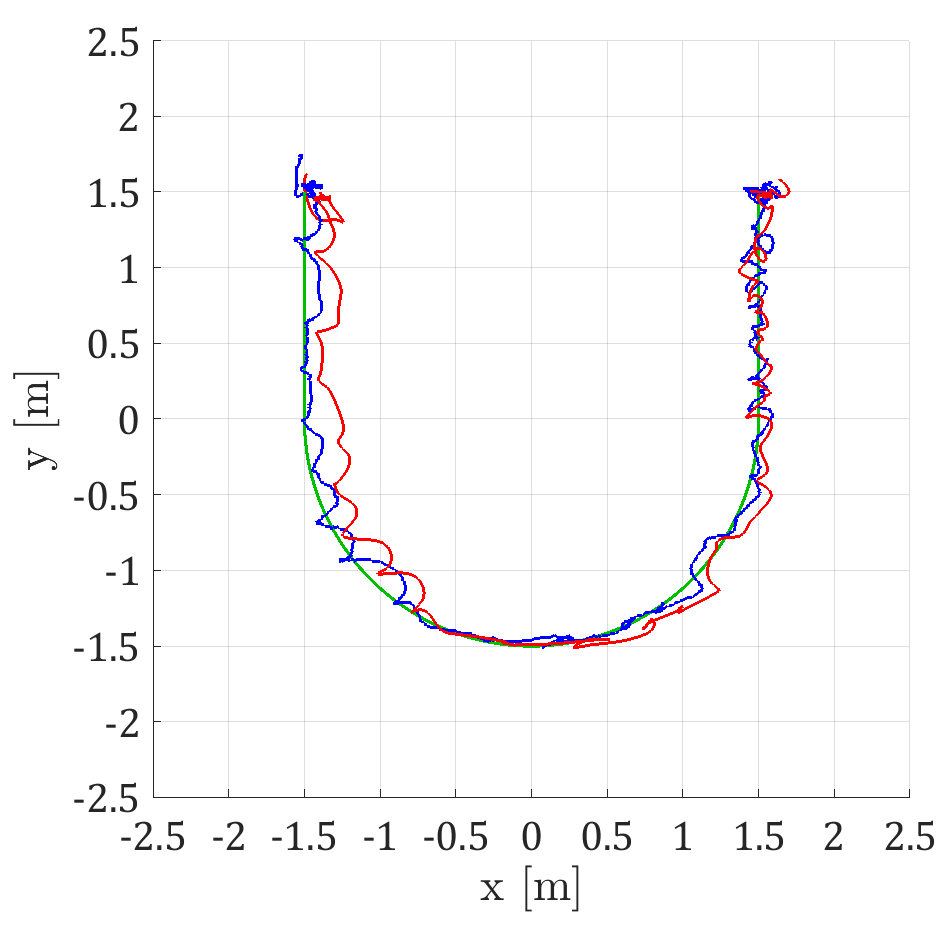}
    \centering
        \includegraphics[width=0.325\textwidth]{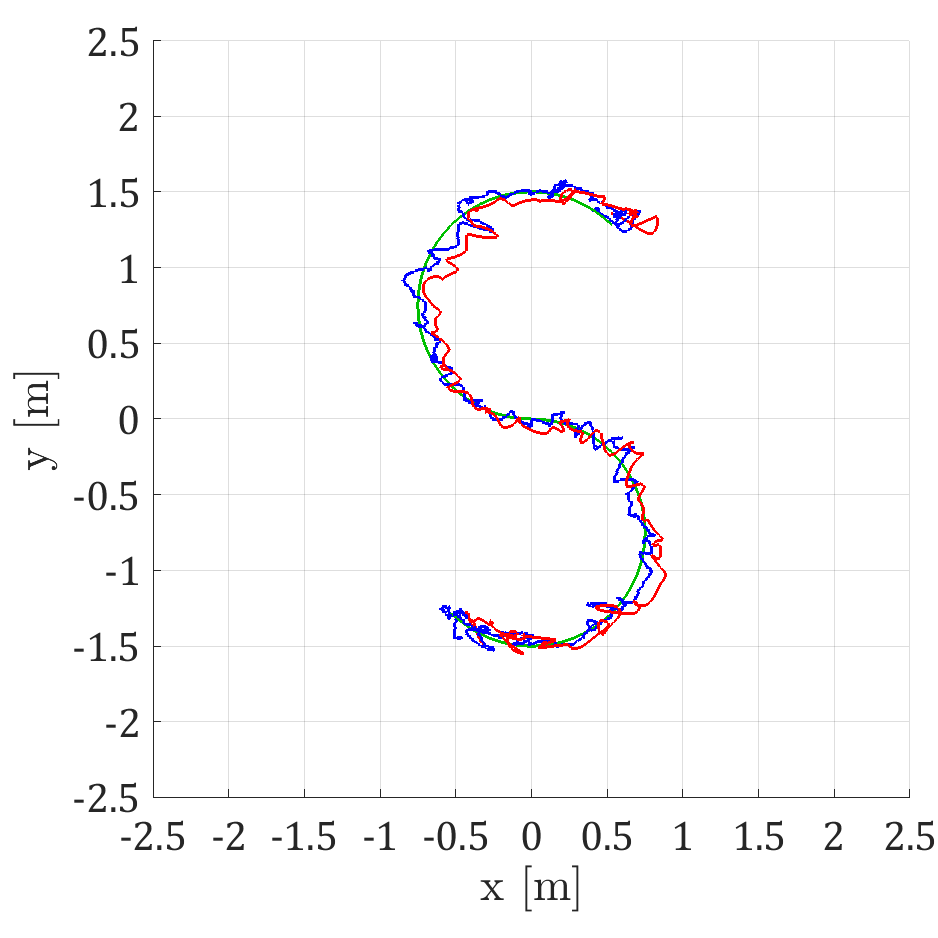}
        \includegraphics[width=0.325\textwidth]{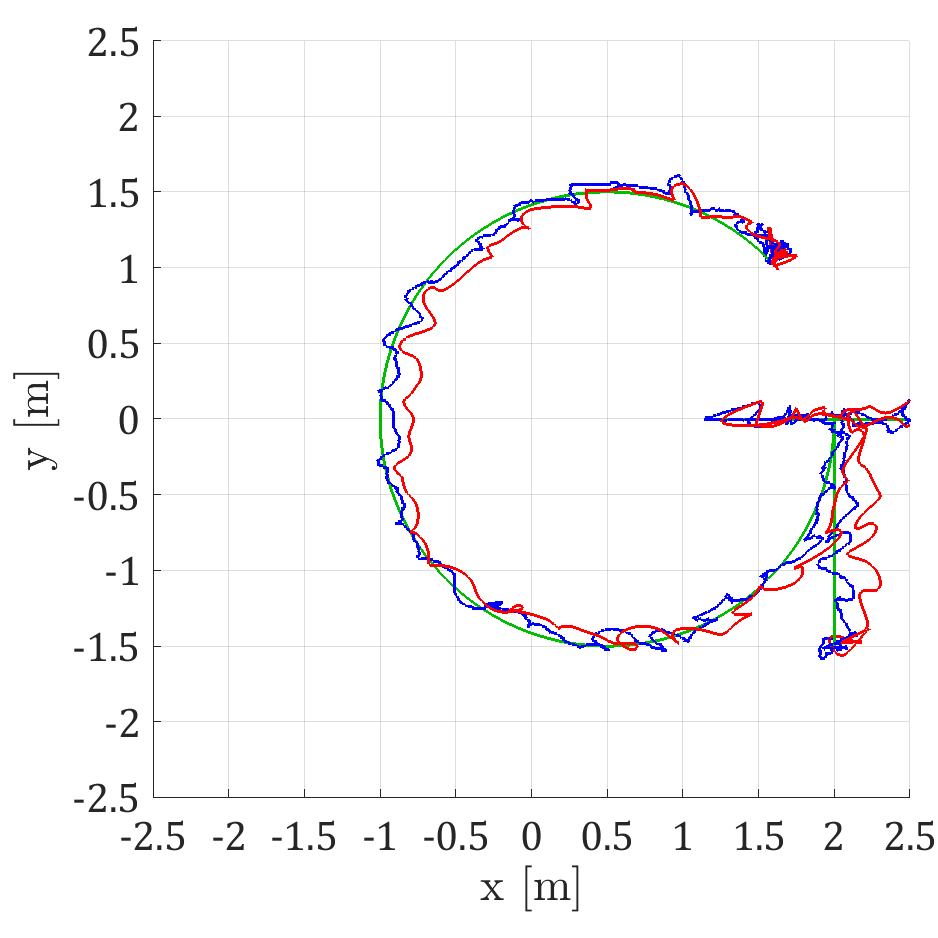}
\caption{Result of static target experiments where MAV follows predefined trajectories with the static responder nodes shifted further away to the new locations $\left(0.369,\ 3.474,\ 1.733 \right)$ and $\left(-0.625,\ 3.461,\ 1.77\right)$ in the frame $F_E$.}
\label{fig:faranc}
\end{figure*}

% \begin{figure*}
%     \centering
%         \includegraphics[width=5.8cm]{exp/path3DN2.png}
%         \includegraphics[width=5.8cm]{exp/path3DT2.png}
%         \includegraphics[width=5.8cm]{exp/path3DU2.png}

%     % \centering
%     %     \includegraphics[width=5.5cm]{exp/path3Dsq_far.png}
%     %     \includegraphics[width=5.5cm]{exp/path3DS2.png}
%     %     \includegraphics[width=5.5cm]{exp/path3DG2.png}

% \caption{Result of static target experiments where MAV follows predefined trajectories with the static responder nodes shifted further away to the new locations $\left(0.369,\ 3.474,\ 1.733 \right)$ and $\left(-0.625,\ 3.461,\ 1.77\right)$ in the frame $F_E$.}
% \label{fig:faranc}
% \end{figure*}

\begin{table}
\centering
\caption{RMSE (\meter) and SD (\meter) of relative position estimate with static anchor at different altitudes.}
\label{tab:nearanc}

\setlength\tabcolsep{5pt}
\begin{tabular}{c || c c c c c c}
\hline\hline
{Altitude}       &$e_{q_x}$      &$e_{q_y}$       &$e_{q_z}$        &$\sigma_{q_x}$ &$\sigma_{q_y}$ & $\sigma_{q_z}$ \\
\hline
$0.6 \mathrm{m}$ &\textbf{0.053} &\textbf{0.094}  &\textbf{0.070}   &0.044  &0.094  &0.025\\
\hline
$0.9 \mathrm{m}$ &\textbf{0.043} &\textbf{0.074}  &\textbf{0.074}   &0.040  &0.074  &0.038\\
\hline
$1.2 \mathrm{m}$ &\textbf{0.054} &\textbf{0.086}  &\textbf{0.139}   &0.049  &0.084  &0.032 \\
\hline\hline
\end{tabular}

\end{table}
\begin{table}
\centering
\caption{RMSE [m] and SD [m] of relative position estimate in the far-anchor experiments.}
\label{tab:faranc}

\setlength\tabcolsep{5pt}
\begin{tabular}{c || c c c c c c}
\hline\hline
{ }       &$e_{q_x}$      &$e_{q_y}$       &$e_{q_z}$        &$\sigma_{q_x}$ &$\sigma_{q_y}$ & $\sigma_{q_z}$ \\
\hline
$1$ &\textbf{0.094} &\textbf{0.040}  &\textbf{0.029}   &0.069  &0.040  &0.025\\
\hline
$2$ &\textbf{0.061} &\textbf{0.044}  &\textbf{0.093}   &0.047  &0.020  &0.016\\
\hline
$3$ &\textbf{0.112} &\textbf{0.068}  &\textbf{0.078}   &0.069  &0.042  &0.022 \\
\hline
$4$ &\textbf{0.099} &\textbf{0.040}  &\textbf{0.074}   &0.041  &0.018  &0.023 \\
\hline
$5$ &\textbf{0.123} &\textbf{0.046}  &\textbf{0.084}   &0.063  &0.046  &0.019 \\
\hline\hline
\end{tabular}

\end{table}

On the MAV's side, there are also two UWB radios but each uses two antennae, thus a total of four requester nodes are presented. The sensor's driver provides complete support to select which antenna to be used in each ranging transaction. Hence, in our setup, we can configure two requester nodes to range to two responder nodes simultaneously using different UWB channels (easily selectable by the sensor's API), then change the antennae to obtain another set of observation. In total, eight distance measurements from every pair of requester and responder can be obtained over a 0.116s long cycle using this switching scheme. We can say that ranging measurements are obtained at a rate of approximately 70Hz. The optical flow data is obtained at 30Hz, IMU data 100Hz and laser range finder data is 40Hz. The target's data sent over UWB signal is configured at 10Hz and VICON data is received at approximately 35Hz. The antennae on the MAV are installed at the corners of a $0.55m \times 0.55m$ square centered around the origin of the frame $F_Q$. The quadcopter is equipped with another embedded computer board, referred to as the high-level board. The computer board has two main tasks. The first task is to organize and collect the UWB range measurement and the second task is to process the camera's image to produce the optical flow data. The high-level board will then send these measurements to a flight control computer (FCC) where the EKF is implemented. The FCC will fuse this information with data from other onboard sensors such as IMU and laser range finder in the EKF and use this estimate in the control loop. All data used for analysis are collected and stored by the high level board including the EKF estimate sent back by the FCC, the ground truth data sent over zigbee from a VICON system and the target's information is sent over the UWB messages.

\subsection{Static target experiments}
In this section we demonstrate the MAV's omni-directional positioning relative to the target. In these so called static target tests, the MAV repeats a trajectory of a $4m\times4m$ square defined in the frame $F_E$ at different altitudes. Fig. \ref{fig:nearanc} shows the results of the test at $0.9m$ altitude. First, the 3D plot of the flight path is presented, then estimates of position, velocity and euler angles compared with ground truth through time are shown in Fig. \ref{fig:nearanc}. All collected data are converted to VICON coordinates system, which is also chosen as the inertial frame of reference $F_E$. The Root Mean Square Error ($e_{q_x}$, $e_{q_y}$, $e_{q_z}$) and Standard Deviation ($\sigma_{q_x}$, $\sigma_{q_y}$, $\sigma_{q_z}$) statistics of the positioning error are shown in Table \ref{tab:nearanc}. Recording of one of these flight tests can be viewed at the begin of the video uploaded to the online link.

\begin{figure*}[h]
    \centering
        \includegraphics[width=0.325\textwidth]{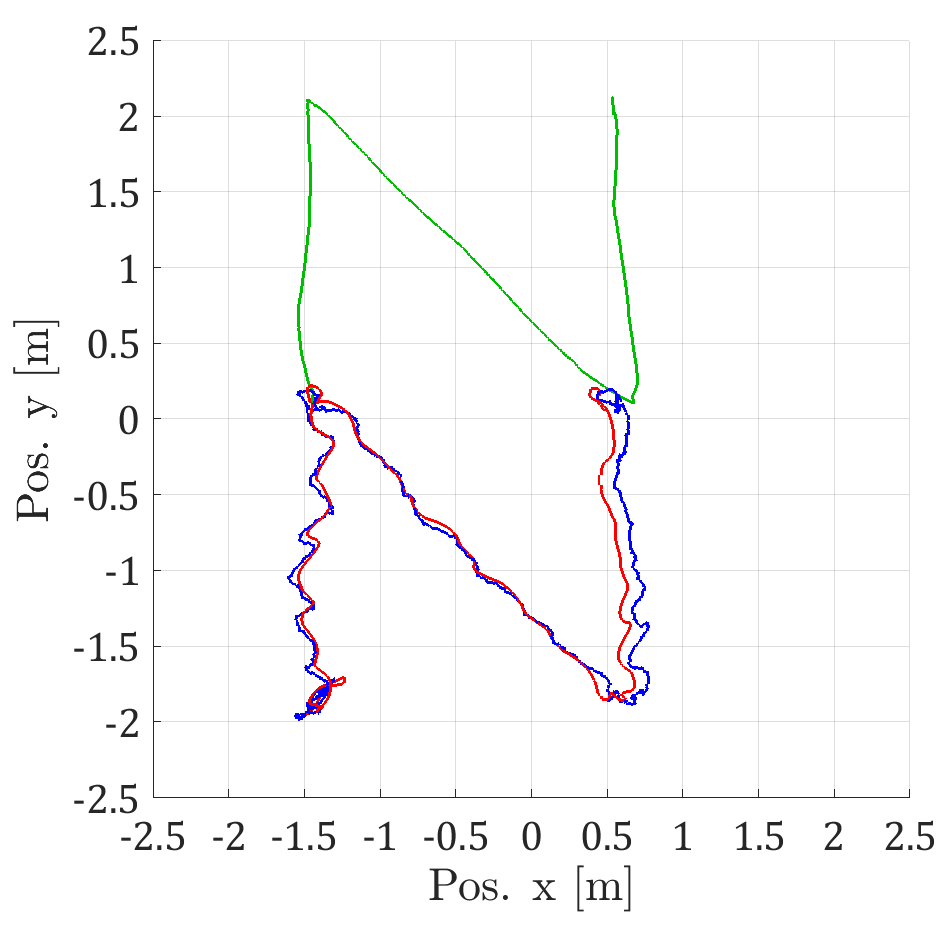}
        \includegraphics[width=0.325\textwidth]{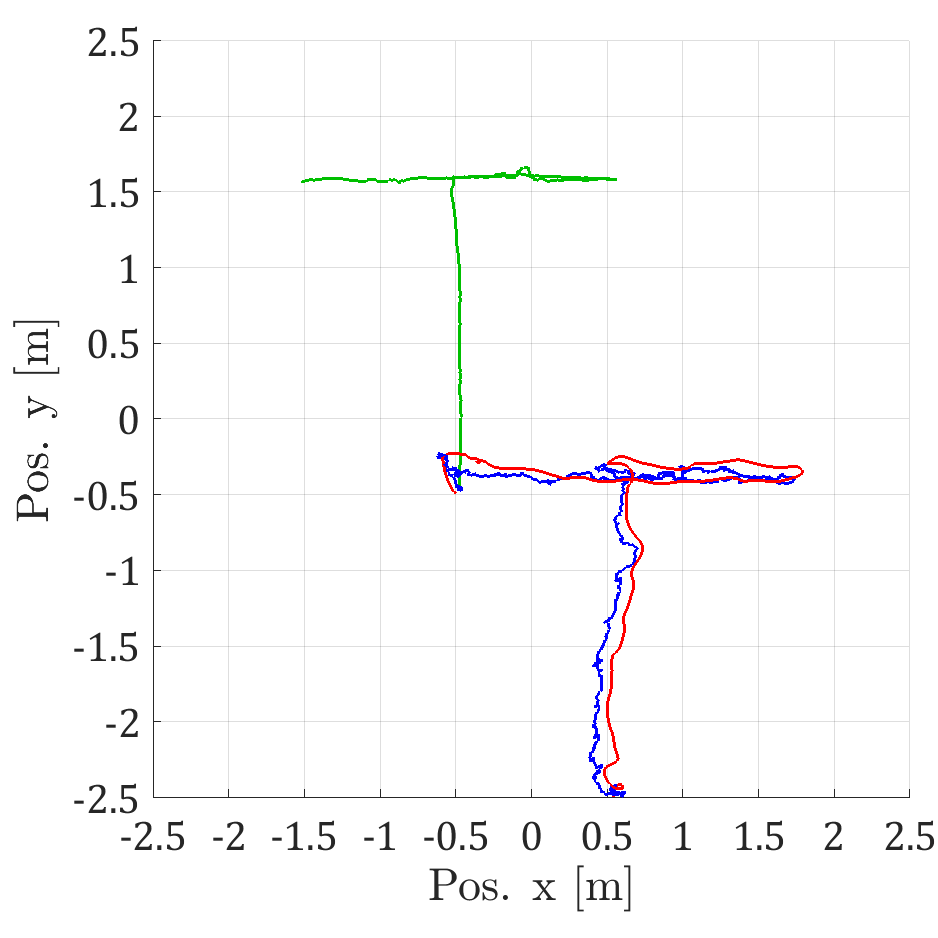}
        \includegraphics[width=0.325\textwidth]{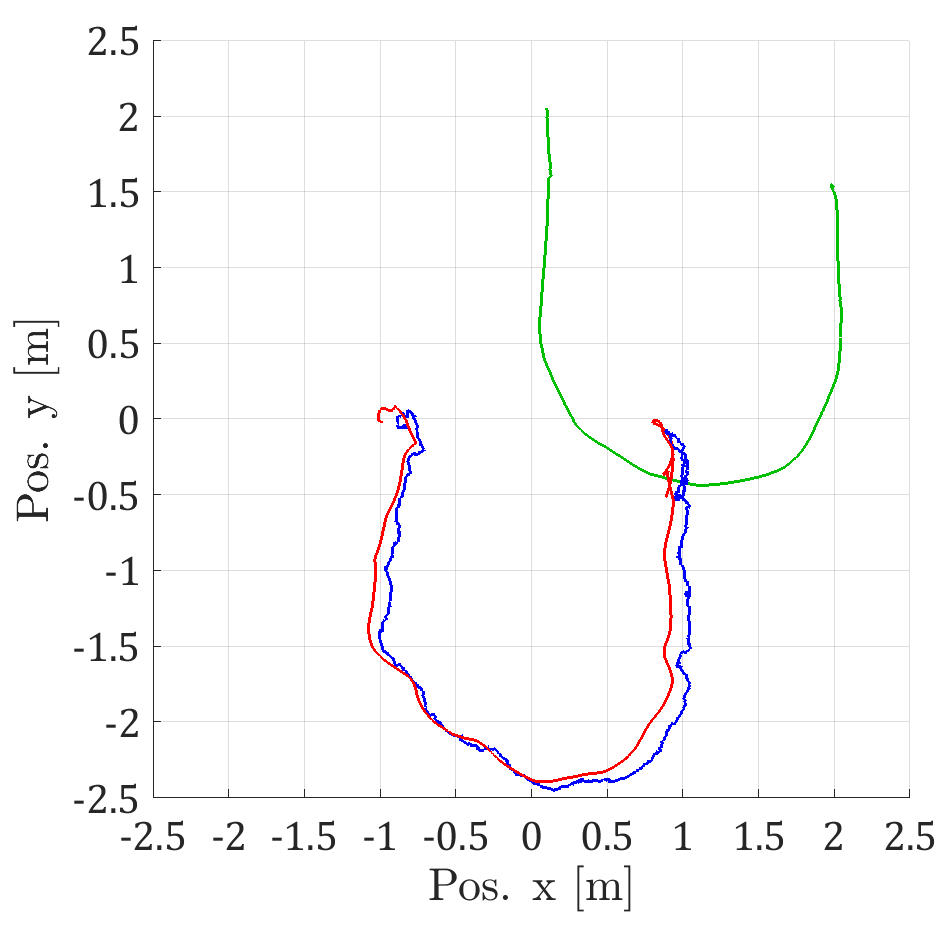}
\caption{Target tracking experiments: The MAV are set to maintain fixed relative position in the frame $F_E$ (a different set relative position is selected in each experiment). The target's path recorded by VICON is plotted in green and the MAV's path is plotted in red. The MAV's relative position estimate is offset by the target's position and displayed in blue. The responders' coordinates in the frame $F_M$ are still separated from each other for about 1m and the values measured by VICON are $\left(-0.019\ 0.700\ 1.428 \right)$ and $\left(-0.012\ -0.338\ 1.419\right)$ respectively. It can be seen that the trajectories made by the MAV are similar to the target's path, which demonstrates the ability of the MAV to robustly estimate and maintain its relative position to the target.}
\label{fig:movanc}
\end{figure*}

% \begin{figure*}[!ht]
%     \centering
%         \includegraphics[width=5.8cm]{exp/movancN.png}
%         \includegraphics[width=5.8cm]{exp/movancT.png}
%         \includegraphics[width=5.8cm]{exp/movancU.png}
% \caption{Target tracking experiments: The MAV are set to maintain fixed relative position to the target (a different set relative position is selected in each experiment). The target's path recorded by VICON is plotted in green and the MAV's path is plotted in red. The MAV's relative position estimate is offset by the target's position and displayed in blue. The responders' coordinates in the frame $F_M$ are still separated from each other for about 1m and the values measured by VICON are $\left(-0.019\ 0.700\ 1.428 \right)$ and $\left(-0.012\ -0.338\ 1.419\right)$ respectively. It can be seen that the trajectories made by the MAV are similar to the target's path, which demonstrates the ability of the MAV to robustly estimate and maintain its relative position to the target.}
% \label{fig:movanc}
% \end{figure*}

Due to space constraints, we cannot carry out the same test where the MAV flies around the responder nodes on a larger square. However we are still interested in finding out to which extent localization is still robust. Thus, we shift the responder nodes approximately 3.5m forward in the $y$ direction of the frame $F_E$. and let the MAVs fly at $0.6m$ altitude. Thus the maximum distance from the MAV to the responders can be up to 5m. Fig. \ref{fig:faranc} shows the paths made the MAVs in these experiments. Similar to the near-anchor tests, the same statistics are done on the relative position estimates and reported in Table \ref{tab:faranc}.

% \begin{table}[h]
% \centering
% \caption{RMSE (\meter) and SD (\meter) of relative position estimate in the far-anchor experiments.}
% \label{tab:faranc}

% \setlength\tabcolsep{5pt}
% \begin{tabular}{c || c c c c c c}
% \hline\hline
% {Experiment}       &$e_{p_x}$      &$e_{p_y}$       &$e_{p_z}$        &$\sigma_{p_x}$ &$\sigma_{p_y}$ & $\sigma_{p_z}$ \\
% \hline
% $1$ &\textbf{0.094} &\textbf{0.040}  &\textbf{0.029}   &0.069  &0.040  &0.025\\
% \hline
% $2$ &\textbf{0.061} &\textbf{0.044}  &\textbf{0.093}   &0.047  &0.020  &0.016\\
% \hline
% $3$ &\textbf{0.112} &\textbf{0.068}  &\textbf{0.078}   &0.069  &0.042  &0.022 \\
% \hline\hline
% \end{tabular}

% \end{table}

These flight tests have demonstrated that our localization system can actually achieve reliable localization data in all directions around the static responder nodes. We also show that our localization is still reliable at a distance up to 5m away from the target with a relatively small spacing of 1m between the responders and $0.55m$ spacing between the requesters. Moreover, the small angle assumption in dealing with correlation flow data is also verified as can be seen in Fig. \ref{fig:nearanc} where the maximum value of the roll and pitch angles are mostly below $5^{o}$ in absolute value.

\subsection{Moving target experiments}

In the first three experiments, so-called \textit{translating target experiments}, the change in the target's orientation is kept relatively small as it moves around. The MAV is set to maintain a fixed position relative to the target. Video recording of these tests can be viewed at the online link. The paths of the target and the MAV are plotted together in Fig. \ref{fig:movanc}.

\begin{figure}
    \centering
        \includegraphics[width=0.45\textwidth]{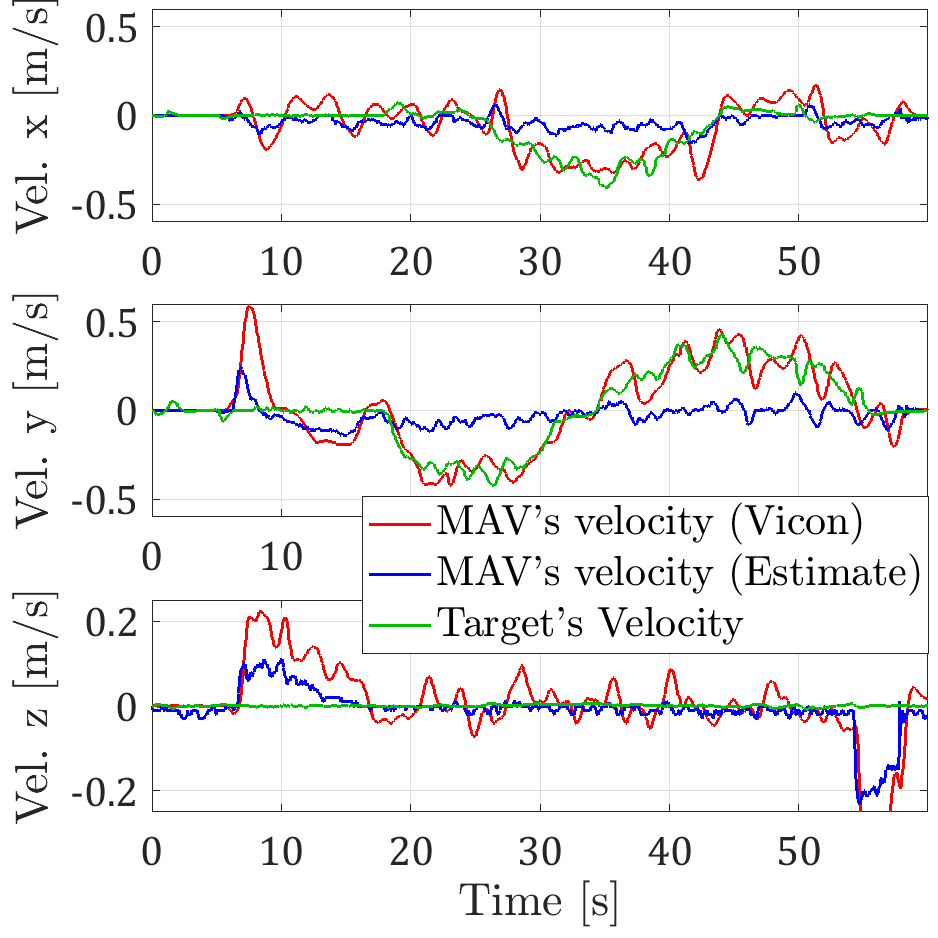}
\caption{Velocities of the target and relative velocity estimate by the MAV in the third experiment. We can see that the target actually moved at 0.4m/s speed. From Fig. \ref{fig:movanc} and the recorded video it can be seen that the MAV can maintain the relative position quite well. This can also be observed in the MAV velocity estimate, which stays nominally close to 0 most of the time.}
\label{fig:movancV}
\end{figure}

\begin{figure}
    \centering
        \includegraphics[width=0.475\textwidth]{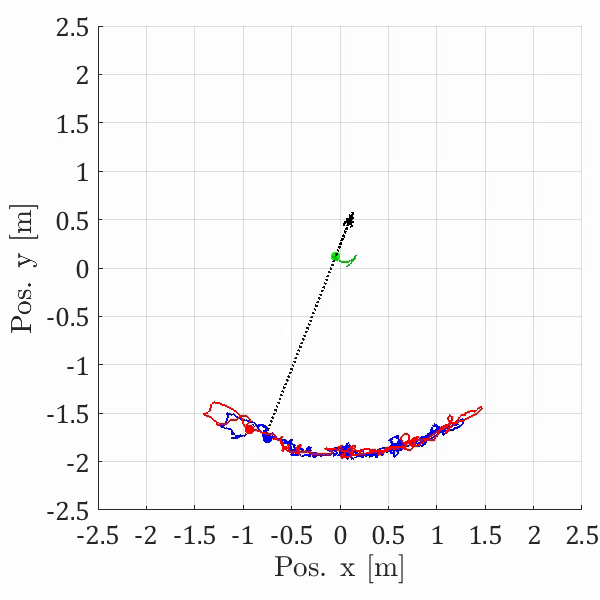}
        \includegraphics[width=0.46\textwidth]{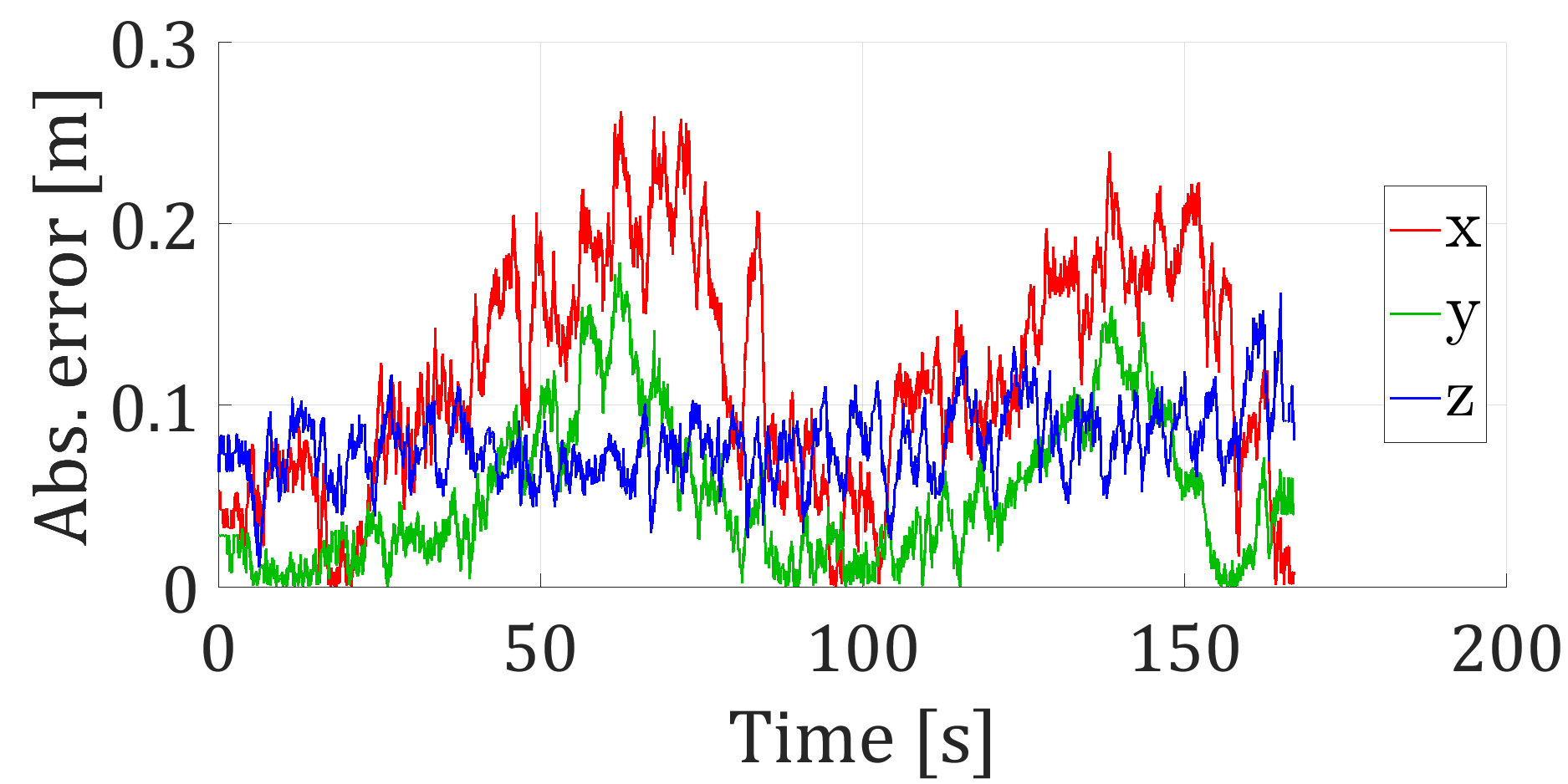}
\caption{Trajectory of the MAV (red, top plot) and its position estimate (blue, top plot) when the set relative position is updated according to the heading of the target. The target and its movement is marked with green, its heading according to its attached IMU is shown by the light blue arrow. The absolute error of relative position estimate in the rotating target experiment over time is shown in the lower plot.}
\label{fig:movancCres}
\end{figure}

In Fig. \ref{fig:movancV}, we show the velocities of the target recorded by VICON along with the MAV's velocity estimate in the third \textit{translating target experiment} as the target moves at highest speed in this test, which is around 0.4m/s in Fig. \ref{fig:movancV}. It can be seen that the velocity estimate does not change very much around zero as the relative position is always maintained.

In the fourth experiment, so-called \textit{rotating target experiment}, a more complicated task is performed where the setpoint is updated by $R^M_E(\psi^M_E)\left[0, -2, 0.75\right]'$. This means that instead of maintaining a fixed position relative to the target, the MAV now has to change its relative position to make sure that it hovers at 2m ``behind" the target's center. This resembles when the MAV has to land on a specific side of the UGV where the landing pad is.

% \begin{figure}
%     \centering
%         \includegraphics[width=0.4\textwidth]{exp/movancCresPE.png}
% \caption{Absolute error of relative position estimate in the rotating target experiment over time.}
% \label{fig:movancCresPE}
% \end{figure}

Table \ref{tab:movanc} summarizes the statistics of the relative position estimation error in these experiments. Fig. \ref{fig:movancCres} shows the absolute error of the relative position estimate of the rotating target experiment as it has the largest RSME. We can see that the maximum error in any direction is approximately 0.25m, thus the absolute 2D localization error of relative positioning in the moving target experiment can be declared as 0.35m.

\begin{table}
\centering
\caption{RMSE (\meter) and SD (\meter) of relative position estimate in moving anchor experiments.}
\label{tab:movanc}
\setlength\tabcolsep{5pt}
\begin{tabular}{c || c c c c c c}
\hline\hline
{Exp.}   &$e_{q_x}$       &$e_{q_y}$        &$e_{q_z}$        &$\sigma_{q_x}$ &$\sigma_{q_y}$ & $\sigma_{q_z}$ \\
\hline
$1$         &\textbf{0.062} &\textbf{0.023}  &\textbf{0.035}   &0.059  &0.018  &0.023\\
\hline
$2$         &\textbf{0.081} &\textbf{0.047}  &\textbf{0.072}   &0.064  &0.039  &0.021\\
\hline
$3$         &\textbf{0.097} &\textbf{0.048}  &\textbf{0.076}   &0.028  &0.020  &0.019 \\
\hline
$4$         &\textbf{0.132} &\textbf{0.065}  &\textbf{0.081}   &0.132  &0.047  &0.020 \\
\hline\hline
\end{tabular}
\end{table}

\section{Conclusion and future works} \label{sec:conclusion}

In this paper we developed a system for relative positioning between a quadcopter and a cooperative target using UWB distance measurements and several other onboard sensors. We show that our use of UWB range measurement can achieve omni-directional relative position estimate that is reliable enough to support feedback-controlled flight, which allows the MAV to autonomously track a moving target. There remain some limitations due to the lack of information regarding the target's velocity. This can certainly be improved in future works by adding a sensor to measure this missing information. Nevertheless we believe the results presented in this paper has consolidated the feasibility of future developments such as precision landing or UGV-MAV collaborative operation, or even multiple MAV formation using the UWB ranging technique.

Another possible direction for future works is to estimate the relative orientation based on UWB measurements. In this way the MAV does not have to receive the IMU data from the target. One can also try to decouple the orientation from other states and only estimate the relative bearing the UGV and MAV to adjust the setpoint accordingly as in the rotating fig: sysarch experiment.

\balance
% \printbibliography
\bibliographystyle{IEEEtran}

\end{document}